\documentstyle{article}
\renewcommand{\theequation}{\thesection.\arabic{equation}}
\newcounter{hran} \renewcommand{\thehran}{\thesection.\arabic{hran}}

\def\bmini{\setcounter{hran}{\value{equation}}
  \refstepcounter{hran}\setcounter{equation}{0}
  \renewcommand{\theequation}{\thehran\alph{equation}}\begin{eqnarray}}

\def\bminiG#1{\setcounter{hran}{\value{equation}}
\refstepcounter{hran}\setcounter{equation}{-1}
\renewcommand{\theequation}{\thehran\alph{equation}}
\refstepcounter{equation}\label{#1}\begin{eqnarray}}

\def\emini{\end{eqnarray}\relax\setcounter{equation}
{\value{hran}}\renewcommand{\theequation}{\thesection.\arabic{equation}}}

\begin{document}
\newcommand{\be}{\begin{equation}}
\newcommand{\ee}{\end{equation}} 
\newcommand{\bea}{\begin{eqnarray}}
\newcommand{\eea}{\end{eqnarray}} 
\newcommand{\nn}{\nonumber}
\newcommand{\bal}{\begin{array}{ll}} 
\newcommand{\eal}{\end{array}}
\newcommand{\la}[1]{\lambda^{#1}}
\newcommand{\Tr}{\rm Tr}
\def\1{{\rm 1 \kern -.10cm I \kern .14cm}} \def\R{{\rm R \kern -.28cm I
\kern .19cm}}
\begin{titlepage} 
\begin{flushright} UFIFT-HEP-97-17\\ RU-97-32  
\\May 1997
\end{flushright} 
\vskip .8cm 
\centerline{\LARGE{\bf {A Model of Yukawa Hierarchies}}} 
\vskip 1.5cm  \centerline{\bf {John K. Elwood, 
Nikolaos Irges and Pierre Ramond}\footnote{Supported in part
by the United States Department of Energy under grants
DE-FG02-97ER41029 and (for PR) DE-FG02-96ER-40559}} 
\vskip .5cm \centerline{\em Institute for
Fundamental Theory,}  \centerline{\em Department of Physics,
University of Florida} \centerline{\em Gainesville FL 32611, USA} 
\vskip 1cm \centerline{\bf {Abstract}}

\noindent 
We present a model for the observed hierarchies among the Yukawa 
couplings of the standard model in the context of an effective low energy 
theory with an anomalous $U(1)$ symmetry. This symmetry, a generic 
feature of superstring compactification, is a remnant of the 
Green-Schwarz anomaly cancellation mechanism. The gauge group is that of 
the standard model, augmented by $X$, the anomalous $U(1)$, and two 
family-dependent phase symmetries $Y^{(1)}$ and $Y^{(2)}$. The correct 
hierarchies are reproduced only when $\sin^2\theta_w=3/8$ at the cut-off.
To cancel anomalies, right-handed neutrinos and other standard model 
singlets must be introduced.  Independently of the charges 
of the right-handed neutrinos, this model produces the same 
neutrino mixing matrix and an {\em inverted} hierarchy of neutrino 
masses. The heaviest is the electron neutrino with a mass $\sim 1$ meV, 
and mixing of the order of $\lambda_c^3$ with each of the other two
neutrinos.  
\end{titlepage}
\section{Introduction} 
Theories of extended objects, superstring theories in particular, 
offer the best hope for resolving the difference between 
General Relativity and Quantum Mechanics. When compactified to four 
dimensions, these models naturally reproduce  Yang-Mills interactions of 
the type found in Nature. Unfortunately, this conceptual matching between 
the $``$fundamental" and the $``$observed" has not yet resulted in a 
detailed 
picture, capable of relating the parameters of the standard model. One 
reason, the disparity of scales, which are $\le $ TeV for the standard
model, 
but $\le 10^{19}$ GeV for superstring theories, can be obviated 
by employing low energy 
supersymmetry.  This allows for a perturbative journey of the standard 
model almost to the Planck scale. There, a more integrated structure 
seems to emerge: the unification of the three gauge couplings, and the 
subject of this paper, the appearance of some order among the Yukawa 
couplings, are both evident.  This order among the Yukawa couplings
can most easily be described as an expansion in the Cabibbo 
angle $\lambda_c$, where we find a geometric {\em interfamily} hierarchy
\begin{equation}
{m_u\over m_t}={\cal O}(\lambda_c^8)\ ;\qquad {m_c\over m_t}={\cal
O}(\lambda_c^4)\ ;  \end{equation}
\begin{equation}{m_d\over m_b}={\cal
O}(\lambda_c^4)\ ; \qquad {m_s\over m_b}={\cal
O}(\lambda_c^2)\ ,\label{eq:l} 
\end{equation}
\begin{equation}
{m_e\over m_\tau}={\cal
O}(\lambda_c^4)\ ; \qquad {m_\mu\over m_\tau}={\cal
O}(\lambda_c^2)\ ,\label{eq:la} 
\end{equation}
among fermions of the same charge. There is also an {\em intrafamily} 
hierarchy
\begin{equation}
{m_b\over m_t}={\cal O}(\lambda_c^3)\ ,
 \qquad {m_b\over m_\tau}={\cal O}(1).\label{eq:lb} 
\end{equation}
 Finally, the CKM quark mixing matrix is of the form~\cite{wolf} 
\be
{\cal U}^{}_{CKM}\sim \left( \begin{array}{ccc}
1&\lambda_c&\lambda_c^3\\
\lambda_c&1&\lambda_c^2\\
\lambda_c^3&\lambda_c^2&1\end{array}  \right)\ ,
\ee
where $\sim$ indicates only order of magnitude estimates. 
Although expressed in terms of fermion mass ratios, these correspond to 
relations among Yukawa couplings. The one exception is $m_b/m_t$, which 
depends on the angle $\beta$ that links the vacuum values of the two 
Higgs in the minimal supersymmetric model. It is therefore possible to 
discuss the origin of these exponents in the context of both exact 
supersymmetry and electroweak symmetries. Below we present a simple model 
that reproduces these hierarchies.

\section{Effective Low Energy Theories with Green-Schwarz Anomalies}
A wide class of superstring theories compactified to four 
dimensions~\cite{orbifold} 
yield $N=1$ supersymmetric effective low energy theories with cut-off 
$M_{\rm string}\le M_{\rm 
Planck}$,  and a universal gauge 
coupling $g_{\rm string}$. Their gauge structure includes a visible sector 
containing at least the standard model gauge groups and several gauged 
phase symmetries $Y^{(1)},Y^{(2)},\dots$. In addition, they contain hidden 
gauge interactions with  unknown gauge structure. These two sectors are 
linked by several gauged Abelian symmetries, one of which, denoted by $X$, 
is anomalous. A subset of these Abelian symmetries are broken {\em below} 
the cut-off by a stringy mechanism~\cite{DSW} that generates a 
Fayet-Iliopoulos term 
in the $D$-term of the $X$ symmetry
\be
\xi^2=-{g^3_{\rm string}\over 192\pi^2}M^2_{\rm Planck}C^{}_{\rm grav}\ 
,\ee
where $C^{}_{\rm grav}$ is the mixed gravitational anomaly of the $X$ 
current
\be
C^{}_{\rm grav}=(X~T~T)\ ,\label{eq:gs0}\ee
the brackets stand for the sum over the particles in the triangle 
graph, and $T$ is the energy momentum tensor. 
For the remainder of this work, we shall denote the cutoff scale
by $M$.  In our convention, $C^{}_{\rm grav}$
 is negative. Since its breaking occurs below the cut-off, it is 
legitimate to include $X$ as a symmetry of the low energy theory. 
We will also require that this vacuum preserve supersymmetry, 
since its scale 
is near the Planck mass.

The anomalies of the $X$-symmetry are compensated at the cut-off by a 
dimension five term in the effective Lagrangian
\begin{equation}{\cal L}={1\over g^2_{\rm string}}\sum_{j} 
k^{}_{j}F^{[j]}_{\mu\nu}
F^{[j]}_{\mu\nu}+i{\eta\over M^{}_{\rm string}}\sum_j k^{}_jF^
{[j]}_{\mu\nu}
{\tilde F}^{[j]}_{\mu\nu}+{\cal L}^{}_{\rm matter}+\cdots\ 
,\label{eq:lc} \end{equation}
where the sum is over the gauge groups, and the $k_j$ are the Kac-Moody 
levels. under the $X$ gauge transformation, the axion-like field $\eta$ 
shifts as a Nambu-Goldstone boson, accounting for the anomalies in the 
$X$ current
\be
\partial_\mu j^X_\mu\sim \sum_j~C^{}_jF_{\mu\nu}^j\widetilde F^j_{\mu\nu}
\ ,\ee
as long as the ratio $C_j/k_j$ is universal. This is the four-dimensional 
equivalent of the Green-Schwarz anomaly cancellation mechanism~\cite{GS}. 
Consistency requires all other 
anomaly coefficients to vanish. 
$C^{}_{\rm color}$, 
$C^{}_{\rm weak}$ and $C^{}_Y$ are the mixed anomalies between the $X$ 
current and the standard model  gauge currents, 
\be  (XG^AG^B)=\delta^{AB}C^{}_{\rm color}\ ;~~~ 
(XW^aW^b)=\delta^{ab}C^{}_{\rm weak}\ ;~~~(XYY)=C^{}_Y\ ,\label{eq:gs1}\ee
where $G^A$ are the QCD currents, and $W^a$ the weak isospin 
currents. We must have
\be
{C^{}_{\rm grav}\over 12}={C^{}_{\rm color}\over k_{\rm color}}={C^{}_{\rm 
weak}\over k_{\rm weak}}={C^{}_{\rm Y}\over k_{\rm Y}}\ne 0\ ,\ee
and  
\be (XY^{(i)}_{}Y^{(j)}_{})=\delta_{}^{ij}C^{(i)}_{}\ .\label{eq:gs2}\ee
All the other anomaly coefficients must vanish by themselves
\be 
(Y^{(i)}_{}Y^{(j)}_{}Y^{(k)}_{})=(Y^{(i)}_{}Y^{(j)}_{}Y)=(Y^{(i)}_{}
G^A_{}G^B_{})=
(Y^{(i)}_{}W^a_{}W^b_{})=(Y^{(i)}_{}YY)=0\ .\ee
as well as 
\be (XYY^{(i)}_{})=(XXY)=(XXY^{(i)}_{})=(Y^{(i)}_{}TT)=0\ .
\label{eq:gs3}\ee
In theories with $N$ symmetries, the number of conditions to be satisfied 
increases as $N^3$, while the number of matter fields is limited by 
asymptotic freedom; it is therefore reasonable to expect that all 
charges could be uniquely determined by anomaly cancellations.

A consequence of this mechanism is that the Weinberg angle at cut-off can 
be understood~\cite{Ib} as a ratio of anomaly coefficients
\be
\tan^2\theta_w={g^2_Y\over g^2_{\rm weak}}={k_{\rm weak}\over 
k_Y}={C_{\rm weak}\over C_Y}\ .\ee
These anomaly coefficients can be computed from the $X$-charges of chiral 
fermions. Such fermions can come in two varieties, those from the 
three chiral 
families and those from standard model pairs with chiral $X$ values. The 
anomaly coefficients from the three chiral families can be related to the 
$X$ charges of the standard model invariants. The minimal supersymmetric 
standard model contains the invariants
 \begin{equation}
{\bf Q}^{}_i{\bf\overline u}^{}_jH^{}_u\ ;
\qquad {\bf Q}^{}_i{\bf\overline d}^{}_jH^{}_d
\ ;\qquad
L^{}_i{\overline e}^{}_jH^{}_d\ ;\qquad 
 H^{}_uH^{}_d\ , \end{equation}
where $i,j$ are the family indices, with $X$ charges 
\be
X^{[u]}_{ij}\ ,\qquad X^{[d]}_{ij}\ ,\qquad X^{[e]}_{ij}\ ,\qquad 
X^{[\mu]}_{}\ ,\ee 
respectively; a simple computation yields
\bea
C^{}_{\rm color}&=&\sum_i^3(X^{[u]}_{ii}+
X^{[d]}_{ii})-3X^{[\mu]}_{}\ ,\\
C^{}_Y+C^{}_{\rm weak}-{8\over 3}C^{}_{\rm color}&=&2\sum_i^3(X^{[e]}_{ii}-
X^{[d]}_{ii})+2X^{[\mu]}_{}\ .\label{eq:anom2}\eea
Since the Kac-Moody levels of the non-Abelian factors are the same, the 
Green-Schwarz condition requires
\be
C^{}_{\rm weak}=C^{}_{\rm color}\ ,\ee 
from which we deduce 
\begin{equation}
C^{}_Y=\sum_i^3({5\over 3} X^{[u]}_{ii}-{1\over
3}X^{[d]}_{ii}+2X^{[e]}_{ii})-3X^{[\mu]}_{}\ .\label{eq:anom3}
\end{equation}
Similar equations hold for the mixed anomalies of the $Y^{(i)}$ currents; 
their vanishing imposes constraints on the $Y^{(i)}$ 
charges of the standard model invariants. 

The further constraint that the Weinberg angle be at its canonical 
$SU(5)$ value,  
$\sin^2\theta_w=3/8$, that is  $3C_Y=5C_{\rm weak}$, yields the 
relations
\begin{equation}
X^{[\mu]}_{}=\sum_i^3(X^{[d]}_{ii}-X^{[e]}_{ii})\ .\label{eq:anom4}
\end{equation} 
 \begin{equation}
C^{}_{\rm color}=\sum_{i}\left[X^{[u]}_{ii}-2X^{[d]}_{ii}+
3X^{[e]}_{ii}\right]
\ ,\label{eq:anom5}
\end{equation}
as well as
 \begin{equation}
C^{}_{\rm color}={1\over 2}\sum_{i\ne j}\left[X^{[u]}_{ij}-2X^
{[d]}_{ij}+3X^{[e]}_{ij}\right]
\ .\label{eq:anom6}\ee
Since $C_{\rm color}$ does not vanish, these equations imply that some 
standard model invariants have non-zero $X$ charges. In the framework of 
an effective field theory, it means that these invariants will appear in 
the superpotential multiplied by fields that balance the excess $X$ 
charge. These higher dimension interactions are suppressed by inverse 
powers of the cut-off~\cite{FN}; this is the origin of 
Yukawa hierarchies and mixings.

A theory with extra Abelian gauged symmetries $X,Y^{(1)},\dots,Y^{(N)}$ 
 will contain $N+1$ standard model singlet chiral superfields $\theta_1,
\dots\theta_{N+1}$, to 
serve as their order parameters. The anomaly-induced 
supersymmetry-preserving vacuum is determined 
by the vanishing of the $N+1$ $D$ terms
\bea
\sum_{a=1}^{N+1} x_a\vert\theta_a\vert^2&=&\xi^2\ ,\\
\sum_{a=1}^{N+1} y^{(k)}_a\vert\theta_a\vert^2&=&0\ ,~~~k=1,2,...,N\ .\eea

These equations can be solved as long as the $(N+1)\times (N+1)$ matrix 
A, with rows equal to the $N+1$ vectors ${\bf x}=(x_1,x_2,...,x_{N+1})$, 
${\bf y}^{(k)}=(y_1^{(k)},y_2^{(k)},...,y^{(k)}_{(N+1)})$ has an 
inverse with a positive first row.   

A typical term in the superpotential, invariant under these $N+1$ 
symmetries will then 
be of the form 
\be
{\bf Q}_i{\bf\overline d}_jH^{}_d\prod_a\left({\theta_a\over 
M}\right)^{n^{[a]}_{ij}}\ ,\label{eq:spot} 
\ee
where holomorphy requires the $n^{[a]}_{ij}$ to be zero or positive 
integers. Invariance under the $N+1$ symmetries yields
\be
X_{ij}^{[d]}+ \sum_a x^{}_a~n_{ij}^{[a]}=0\ ,\ee
  \be
Y_{ij}^{(k)~[d]}+ \sum_a y^{(k)}_a~n_{ij}^{[a]}=0\ ,~~~k=1,2,...,N\ .\ee
These involve the same matrix $A$, and here a solution also requires that 
$\det A\ne 0$, linking hierarchy to vacuum structure. Evaluated at the 
vacuum values of the $\theta_a$ fields, the terms shown above can produce 
a family-dependent Yukawa hierarchy.

A successful model of this type is highly constrained: it must satisfy 
all anomaly conditions and reproduce the observed Yukawa hierarchies.
Additionally, the 
breaking triggered by the anomalous $U(1)_X$ must preserve supersymmetry, 
as well as the standard model gauge symmetries. In searching for models 
of this type, we assume that the $X$ charge is 
family-independent, and that the  $Y^{(i)}$ charges are traceless over 
the families. In this way, the $Y^{(i)}$ are responsible for the 
interfamily hierarchy and mixing while the $X$ charges account for the 
intrafamily structure.

The  role of the   anomalous symmetry in generating hierarchies
has been proposed  earlier~\cite{IR,BR,NIR,JS,CCK}, but with  a
family-dependent $X$ symmetry.
 In previous works, it was pointed out how the Weinberg angle is related 
to the hierarchy~\cite{BR,NIR} and that the seesaw mechanism 
implies $R$-parity 
conservation~\cite{BILR}.  Below we present a model in which all of these 
features are satisfied.

\section{The Model} 
In this simple illustrative model, there are three 
 gauged symmetries beyond the standard model: a family-independent 
anomalous $X$, and two family-traceless symmetries 
$Y^{(1)},Y^{(2)}$. On the three chiral families, they are
\be
Y^{(1)}=(B+L) \left( \begin{array}{ccc}
2&0&0\\ 0&-1&0\\ 0&0&-1
\end{array}  \right)\ ,\label{eq:yon}\ee
where $B$ and $L$ are baryon and lepton numbers, and the diagonal matrix 
is in family space. The other charges are
\be
Y^{(2)}= \left( \begin{array}{ccc}
1&0&0\\ 0&0&0\\ 0&0&-1
\end{array}  \right)\ ,\label{eq:yone}\ee
for ${\bf Q},{\bf\overline u},\overline e$ and zero for $L,{\bf\overline 
d}$. We assume that the only dimension-three term in the superpotential 
is the 
Yukawa coupling for the top quark, 
\be
W={\bf Q}_3{\bf\overline u}_3H_u\ .\ee
The family tracelessness of $Y^{(1)}$ and $Y^{(2)}$ insures the vanishing 
of the contribution of the the chiral families of many anomaly 
coefficients
\be
(Y^{(i)}_{}G^A_{}G^B_{})_f=(Y^{(i)}_{}W^a_{}W^b_{})_f=(Y^{(i)}_{}YY)_f
=(Y^{(i)}_{}TT)=(XYY^{}_i)_f=0\ .\ee
The model assumes no fermions with standard model quantum numbers except 
for those in the MSSM. It therefore follows from the above equations that 
the Higgs pair is vector-like with respect to the $Y^{(1,2)}$ charges. 
Since $(XYY^{(1,2)})$ must vanish over the Higgs pair, we infer that their 
charges are also vector-like with respect to $X$. Hence all the charges of 
the $\mu$ term vanish
\be
X^{[\mu]}=Y^{(1)~[\mu]}=Y^{(2)~[\mu]}=0\ ,\ee
which is also favored by the independent vacuum analysis~\cite{BILR}. 
The other anomaly conditions involving the hypercharge must be satisfied 
by the chiral fermions
\be
(Y^{(1)}_{}Y^{(1)}_{}Y)_f=(Y^{(2)}_{}Y^{(2)}_{}Y)_f
=(Y^{(1)}_{}Y^{(2)}_{}Y)_f=0 \ .\ee
For these to hold, it is not sufficient to invoke family-tracelessness, 
but our assignment clearly satisfies these equations. Other anomaly 
conditions that do not involve standard model currents are computed to be
\be
(Y^{(1)}_{}Y^{(1)}_{}Y^{(1)}_{})_f=(Y^{(1)}_{}Y^{(1)}_{}Y^{(2)}_{})_f=6\ 
,\ee
\be
(Y^{(2)}_{}Y^{(2)}_{}Y^{(2)}_{})_f=(Y^{(1)}_{}Y^{(2)}_{}Y^{(2)}_{})_f=0\ .
\ee
These anomalies need to be canceled by other fields, 
some of which must be the
$\theta$ fields whose vacuum values 
break the $X$ and $Y^{(1,2)}$ symmetries. These do not 
suffice to saturate the anomaly conditions with rational 
charge assignments, however.  More fields must  
be added;  some will be interpreted as the 
right-handed partners of the standard model neutrinos.

The charges of the $\theta$ fields are constrained by the 
observed Yukawa hierarchies, which are reproduced by
\be
A^{-1}= \left( \begin{array}{ccc}
1&0&0\\ 1&0&-1\\ 1&1&-1
\end{array}  \right)\ ,
\ee
so that all  three $\theta$ fields ave the same vacuum expectation value
\be
\vert<\theta_1>\vert=\vert<\theta_2>\vert=\vert<\theta_3>\vert=
\xi \ .\ee
Their charges are given by
\be
A= \left( \begin{array}{ccc}
1&0&0\\ 0&-1&1\\ 1&-1&0
\end{array}  \right)\ ,
\ee
and the $\theta$ sector contributions to the anomalies are
\be
(Y^{(1,2)}_{}TT)_\theta=
(Y^{(1)}_{}Y^{(1)}_{}Y^{(1)}_{})_\theta
=(Y^{(2)}_{}Y^{(2)}_{}Y^{(2)}_{})_\theta=0\ ,\ee
\be
(Y^{(1)}_{}Y^{(1)}_{}Y^{(2)}_{})_\theta 
=(Y^{(1)}_{}Y^{(2)}_{}Y^{(2)}_{})_\theta=-1\ .
\ee
Clearly more fields must be added, and from hereon, our construction is 
somewhat arbitrary, guided mostly by anomaly cancellation with 
rational charges. As an example, we introduce three $SO(10)$-like 
right-handed neutrinos 
with $Y^{(1,2)}$ charges of the same family structure as the chiral 
families:
\begin{center}
\begin{tabular}{|c|c|c|c|}
           \hline & & & \\
$~{\rm Charge}~$
&$~\overline N_1~$&$~\overline N_2~$&$~\overline N_3~$\\
& & & \\ \hline 
\hline   & & & \\
$~X~$
&$~-1/2~$&$-1/2~$&$~-1/2~$\\& & & \\
\hline    & & & \\  
$Y^{(1)}~$
&$~-2~$&$~1~$&$~1~$\\ & & & \\
\hline   & & & \\
$Y^{(2)}~$
&$~-1~$&$~0~$&$~~1~$\\  & & & \\
\hline
\end{tabular} \end{center}
\vspace{0.4cm}
which contribute to three anomaly coefficients
\be
(Y^{(1)}_{}Y^{(1)}_{}Y^{(1)}_{})_{\overline N} 
=-6\ ;\ \ (Y^{(1)}_{}Y^{(1)}_{}Y^{(2)}_{})_{\overline N}=-3
\ ;\ \ (Y^{(1)}_{}Y^{(2)}_{}Y^{(2)}_{})_{\overline N}=-1\ .
\ee
Their $X$ charges insure, through the seesaw mechanism, $R$-parity
conservation~\cite{BILR}. We also introduce four additional 
standard model singlets to cancel the remaining anomalies:

\begin{center}
\begin{tabular}{|c|c|c|c|c|}
           \hline & & & &\\
$~{\rm Charge}~$
&$~S_1~$&$~S_2~$&$~S_3~$&$~S_4~$\\
& & & &\\ \hline 
\hline   & & & &\\
$~X~$
&$~1~$&$~0~$&$~0~$&$~-1~$\\& & & &\\
\hline    & & & &\\  
$Y^{(1)}~$
&$~-3/2~$&$~-1/2~$&$~1/2~$&$~3/2~$\\ & & & &\\
\hline   & & & &\\
$Y^{(2)}~$
&$~1/2~$&$~3/2~$&$~-1/2~$&$~-3/2~$\\  & & & &\\
\hline
\end{tabular} \end{center}
\vspace{0.4cm}
Their contributions to the anomalies are
\be
(Y^{(1)}_{}Y^{(1)}_{}Y^{(1)}_{})_{S} 
=0\ ;\qquad (Y^{(1)}_{}Y^{(1)}_{}Y^{(2)}_{})_{S}=-2\ ;\ee
\be
(Y^{(1)}_{}Y^{(2)}_{}Y^{(2)}_{})_{S}=2\ ;\qquad 
(Y^{(2)}_{}Y^{(2)}_{}Y^{(2)}_{})_{S}=0\ .\ee
Hence the sum of all the $(Y^{(i)}Y^{(j)}Y^{(k)})$ coefficients 
vanish.  It is worth pointing out that the structure of the
$S$-field sector is not crucial to any of the predictions of this 
paper, and its inclusion serves only to demonstrate a particular mechanism
for canceling anomalies.  In fact, one must generally take care not to
unleash unwanted vacuum flat directions as more $S$ fields are 
included in the model.  

Three of the five $X$ charges of the chiral families are determined from 
the conditions $(XXY)_f=0$, $C_{\rm color}=C_{\rm weak}$, and 
$\sin^2\theta_w=3/8$ at unification, yielding
\be
X_{\bf Q}=X_{\overline{\bf u}}=X_{\overline e}\equiv a
\ ;\qquad 
X_{L}=X_{\overline{\bf d}}\equiv d\ .\ee
Using the top quark Yukawa constraint and the neutrality of the 
$\mu$ term,we find
\be
C_{\rm color}=C_{\rm weak}={3\over 5}C_Y=3a+d\ .\ee
We also find 
\be
(XY^{(1)}Y^{(2)})_f=0 .\ee
Hence 
\be
(XY^{(1)}Y^{(2)})=(XY^{(1)}Y^{(2)})_{\overline N}
+(XY^{(1)}Y^{(2)})_S=0\ .\ee
The remaining anomalies that do not vanish are all determined 
in terms of $a$ and $d$
\be
(XXX)=10a^3+5d^3+{21\over 8}\ ;\ \ C_{\rm grav}=10a+5d+{3\over 
2} ;\ee
\be
(XY^{(1)}Y^{(1)})=12a+14d+{5\over 2}\ ;\qquad 
(XY^{(2)}Y^{(2)})=20a+{5\over 2}\ .\ee
The expansion parameter is therefore fully determined in terms of 
the string coupling constant, the Planck mass and the 
$X$ charges of the chiral families.

\subsection{Quark Yukawa Hierarchies}
The family structure of the charge $2/3$ Yukawa couplings is 
determined by the $Y^{(1,2)}$  charges as well as by 
$A^{-1}$. The invariant Yukawa interaction in the 
superpotential is
\be {\bf Q}^{}_i\bar{\bf u}^{}_jH^{}_u
{\bigl ( {\theta_1 \over M} \bigr )}^{n^{(1)}_{ij}}
{\bigl ( {\theta_2 \over M} \bigr )}^{n^{(2)}_{ij}}
{\bigl ( {\theta_3 \over M} \bigr )}^{n^{(3)}_{ij}}
\ ,\label{eq:uterm}\ee
Invariance under the three charges yields
\bea
n^{(1)}_{ij}&=&0\ ,\\
n^{(2)}_{ij}&=&Y^{(2)~[u]}_{ij}\ ,\\
n^{(3)}_{ij}&=&-Y^{(1)~[u]}_{ij}+Y^{(2)~[u]}_{ij}\ ,
\eea
where we have used the fact that $X$ is family independent and 
that $X^{[u]}=0$ from the top quark mass. The exponents 
$n^{(1)}_{ij},n^{(2)}_{ij}$ are easily computed;  all are 
either zero or positive integers, so that there are no 
supersymmetric zeros \cite{LENS}. The orders of magnitude in the charge 
$2/3$ Yukawa matrix are therefore
 \be
Y_{}^{[u]}\sim\left( \begin{array}{ccc}
\lambda^8 &\lambda^5&\lambda^3\\ \lambda^7&\lambda^4&\lambda^2\\
\lambda^5&\lambda^2&1\end{array}  \right)\ ,\ee
where 
\be
\lambda={\vert\theta\vert\over M}\ .\ee
This matrix reproduces the geometric interfamily hierarchy in this sector, 
\be
{m_u\over m_t}\sim \lambda_c^8\ ,\qquad {m_c\over m_t}\sim
\lambda_c^4\ ,\ee
and its left-handed diagonalization is of the CKM form
\be
\left( \begin{array}{ccc}
1 &\lambda&\lambda^3\\ \lambda&1&\lambda^2\\
\lambda^3&\lambda^2&1\end{array}  \right)\ ,\ee
with the expansion parameter identified with the Cabibbo angle 
$\lambda_c$.

In the down sector, the corresponding exponents are given by 
\bea
p^{(1)}_{ij}&=&-X^{[d]}\ ,\\
p^{(2)}_{ij}&=&-X^{[d]}+Y^{(2)~[d]}_{ij}\ ,\\
p^{(3)}_{ij}&=&-X^{[d]}-Y^{(1)~[d]}_{ij}+Y^{(2)~[d]}_{ij}\ .
\eea
To avoid supersymmetric zeros in all matrix 
elements, $X^{[d]}$ must be a negative integer or zero. The 
$Y^{(1,2)}$ charges of the down matrix elements are
\be 
(Y_{}^{(1)~[d]},Y_{}^{(2)~[d]})=\left( \begin{array}{ccc}
(~~0,-1) &(1,-1)&(1,-1)\\ (-1,-2)&(0,-2)&(0,-2)\\
(-1,-3)&(0,-3)&(0,-3)\end{array}  \right)\ .\ee
A supersymmetric zero can develop in the $(33)$ position if 
$X^{[d]}> -3$. To reproduce the observed hierarchy we must 
therefore require
\be
X^{[d]}\le -3\ .\ee
With this proviso, the down Yukawa matrix orders of magnitude 
are
\be 
Y_{}^{[d]}\sim\lambda_{c}^{-3X^{[d]}-6}\left( \begin{array}{ccc}
\lambda_c^{4} &\lambda_c^{3}&\lambda_c^{3}\\ 
\lambda_c^{3}&\lambda_c^{2}&\lambda_c^{2}\\
\lambda_c^{}&1&1\end{array}  \right)\ ,\ee
which leads to its left-diagonalization by  a matrix with the 
CKM structure. Hence the CKM mixing matrix is reproduced, with 
the correct identification of the expansion parameter with the 
Cabibbo angle
\be
{\cal U}^{}_{CKM}\sim\left( \begin{array}{ccc}
1 &\lambda_c^{}&\lambda_c^{3}\\ \lambda_c^{}&1&\lambda_c^{2}\\
\lambda_c^{3}&\lambda_c^{2}&1\end{array}  \right)\ ,\ee
and the hierarchy
\be
{m_d\over m_b}\sim\lambda_c^4\ ,\qquad {m_s\over m_b}\sim \lambda_c^2\
,\ee
All interfamily hierarchies are reproduced in the quark 
sectors. The intrafamily quark hierarchy is 
\be
{m_b\over m_t}= \cot\beta\lambda_{c}^{-3X^{[d]}-6}\ ,\ee
implying a suppression determined by the value of the color 
anomaly. 

\subsection{Charged Lepton Hierarchies}
The exponents of the charged lepton sector are given by
\bea
q^{(1)}_{ij}&=&-X^{[e]}\ ,\\
q^{(2)}_{ij}&=&-X^{[e]}+Y^{(2)~[e]}_{ij}\ ,\\
q^{(3)}_{ij}&=&-X^{[e]}-Y^{(1)~[e]}_{ij}+Y^{(2)~[e]}_{ij}\ ,
\eea
indicating that $X^{[e]}$ must also be a negative integer or 
zero. This is consistent with 
the canonical value of the Weinberg angle, for which  
$X^{[e]}=X^{[d]}$. The $Y^{(1,2)}$ charges in the charged lepton 
Yukawa matrix are
\be
(Y_{}^{(1)~[e]},Y_{}^{(2)~[e]})=\left( \begin{array}{ccc}
(~~0,-1) &(3,-2)&(3,-3)\\ 
(-3,-1)&(0,-2)&(0,-3)\\
(-3,-1)&(0,-2)&(0,-3)\end{array}  \right)\ .\ee
If $X^{[e]}\le -6$ there are no supersymmetric zeros, and one 
can check that the observed $e-\mu-\tau$ hierarchy is not 
reproduced. If $X^{[e]}= -5$, there is one supersymmetric 
zero in the $(13)$ position, but again the hierarchy comes out 
wrong. It is only with $X^{[e]}\ge -4$, with at least two supersymmetric 
zeros in the $(12)$ and $(13)$ positions, that one reproduces the 
observed pattern with  
\be 
Y_{}^{[e]}\sim\lambda_{c}^{-3X^{[e]}-6}\left( \begin{array}{ccc}
\lambda_c^{4} &0&0\\ 
\lambda_c^{7}&\lambda_c^{2}&1\\
\lambda_c^{7}&\lambda_c^2&1\end{array}  \right)\ .\ee
Thus the constraints
\be 
-3\ge X^{[d]}=X^{[e]}\ge -4\ ,\ee
reproduce the lepton hierarchy
\be
{m_e\over m_\tau}\sim\lambda_c^4\ ,\qquad {m_\mu\over m_\tau}\sim
\lambda_c^2\ ,\ee
with  two solutions:
\be
{m_b\over m_\tau}\sim 1\ ;\qquad {m_b\over m_t}\sim 
\cot\beta\lambda_c^3~~~~{\rm or} ~~~~\cot\beta\lambda_c^6\ .\ee
The latter case is not viable as it implies that $\beta\sim 
0$, but the first yields an acceptable mass ratio with 
$\tan\beta\sim 1$. In either case, this ratio is naturally 
suppressed. The left-diagonalization of this matrix yields half 
the lepton mixing matrix
\be
\left( \begin{array}{ccc}
1&\lambda_c^9&\lambda_c^{11}\\ 
\lambda_c^{9}&1&1\\
\lambda_c^{11}&1&1\end{array}  \right)\ ,\ee
indicating large mixing between the $\mu$ and the $\tau$, and no mixing
with the electron.

\subsection{Neutrino Hierarchies and Mixing}
The coupling of the right-handed neutrinos to the standard model is
of the form
\be
L_i{\overline 
N_j}H^{}_u{\bigl ( {\theta_1 \over M} \bigr )}^{r^{(1)}_{ij}}
{\bigl ( {\theta_2 \over M} \bigr )}^{r^{(2)}_{ij}}
{\bigl ( {\theta_3 \over M} \bigr )}^{r^{(3)}_{ij}}\
,\ee
with the integer exponents given by
\bea
r^{(1)}_{ij}&=&-X^{[\nu]}\ ,\\
r^{(2)}_{ij}&=&-X^{[\nu]}+Y^{(2)~[\nu]}_{ij}\ ,\\
r^{(3)}_{ij}&=&-X^{[\nu]}-Y^{(1)~[\nu]}_{ij}+Y^{(2)~[\nu]}_{ij}\ ,
\eea
and
\be
X^{[\nu]}=d-2a-{1\over 2}\ ,\ee
must be a negative integer or zero to avoid supersymmetric zeros
everywhere.  $Y^{(i)~[\nu]}$ are the charges of the invariants $L_i\overline
N_jH_u$ given by 
\be
(Y_{}^{(1)~[\nu]},Y_{}^{(2)~[\nu]})=\left( \begin{array}{ccc}
(~~0,1) &(3,2)&(3,3)\\ 
(-3,1)&(0,2)&(0,3)\\
(-3,1)&(0,2)&(0,3)\end{array}  \right)\ .\ee
If $X^{[\nu]}=0$, there is a supersymmetric zero in the $(12)$
entry. If $X^{[\nu]}\le -1$ and integer, there are no supersymmetric
zeros, and if $X^{[\nu]}$ is positive or fractional there are no
couplings between the standard model and the $\overline N$.

First if $X^{[\nu]}\le -1$, we have
\be 
Y_{}^{[\nu]}\sim\lambda_{c}^{-3X^{[\nu]}}\left( \begin{array}{ccc}
\lambda_c^{2} &\lambda_c^{}&\lambda_c^{3}\\ 
\lambda_c^{5}&\lambda_c^{4}&\lambda_c^6\\
\lambda_c^{5}&\lambda_c^4&\lambda_c^6\end{array}  \right)\ .\ee
On the other hand, if $X^{[\nu]}=0$, the same matrix reads
\be 
Y_{}^{[\nu]}\sim\left( \begin{array}{ccc}
\lambda_c^{2}&0 &\lambda_c^3\\ 
\lambda_c^{5}&\lambda_c^{4}&\lambda_c^6\\
\lambda_c^{5}&\lambda_c^4&\lambda_c^6\end{array}  \right)\ .\ee

Additionally, Majorana mass terms for the right-handed neutrinos,

\be
M{\overline N_i}{\overline 
N_j}{\bigl ( {\theta_1 \over M} \bigr )}^{t^{(1)}_{ij}}
{\bigl ( {\theta_2 \over M} \bigr )}^{t^{(2)}_{ij}}
{\bigl ( {\theta_3 \over M} \bigr )}^{t^{(3)}_{ij}}\
,\ee
are generally allowed, where the powers $t_{ij}^{(1,2,3)}$ are given by
Eqs. (3.79-81) with $X^{[\nu]}=-{1 \over 2}$.
The charges of the $\overline N_i\overline N_j$
combinations are
\be
(Y_{}^{(1)~[0]},Y_{}^{(2)~[0]})=\left( \begin{array}{ccc}
(-4,-2) &(-1,-1)&(-1,0)\\ 
(-1,-1)&(2,0)&(2,1)\\
(-1,0)&(2,1)&(2,2)\end{array}  \right)\ .\ee
which implies supersymmetric zeros in the $(11)$ and $(22)$ positions,
and the Majorana mass matrix
\be 
Y^{[0]}\sim\lambda_{c}^2\left( \begin{array}{ccc}
0&1&\lambda_c^{2} \\ 
1&0&\lambda_c^{}\\
\lambda_c^2&\lambda_c^{}&\lambda_c^3\end{array}  \right)\ .\ee
Diagonalization of this matrix yields the eigenvalues 
$\lambda_c^2$, $\lambda_c^2$, and $\lambda_c^5$, which describe,  
in the absence of electroweak breaking, one  Dirac pair with  
mass $\sim M\lambda_c^2$ and one Majorana mass 
$\sim M\lambda_c^5$.

Electroweak breaking causes these states to mix with the 
neutrinos of the standard model, through the seesaw 
mechanism~\cite{SEESAW}. Two cases must be considered separately.

When $X^{[\nu]}\le -1$, the effective neutrino Yukawa mixing is 
given by
\be 
\widehat Y^{[\nu]}\sim{v_u^2\over M}\lambda_{c}^{-6X^{[\nu]}}
\left( \begin{array}{ccc}
1&\lambda_c^{3} &\lambda_c^3\\ 
\lambda_c^{3}&\lambda_c^{6}&\lambda_c^{6}\\
\lambda_c^3&\lambda_c^{6}&\lambda_c^6\end{array}  \right)\ .\ee
This leads to the neutrino masses
\be
m_{\nu_e}\sim {v_u^2\over M}\lambda_{c}^{-6X^{[\nu]}}\ ;\qquad
m_{\nu_\mu}\sim m_{\nu_\tau}\sim {v_u^2\over 
M}\lambda_{c}^{-6X^{[\nu]}+6}\ .\ee
The MNS~\cite{MNS} neutrino mixing matrix works out to be
\be
{\cal U}_{MNS}\sim
\left( \begin{array}{ccc}
1&\lambda_c^{3} &\lambda_c^3\\ 
\lambda_c^{3}&1&1\\
\lambda_c^3&1&1\end{array}  \right)\ ,\ee 
which shows that the electron neutrino mixing angle with the 
others is of the order of $\lambda_c^3$ while the $\mu$ and 
$\tau$ neutrinos mix together with large angles. With $v_u\sim 
250$ GeV and $M\sim 10^{17}$ GeV, the electron neutrino mass is at 
most 
\be
m_{\nu_e}\sim \lambda_c^6\times~(.6~{\rm meV})\ ,\ee
when $X^{[\nu]}=-1$. These values for the neutrino masses are far
too small for the MSW effect to operate within the sun \cite{MSW}.
Neither will vacuum oscillations explain the solar neutrino problem,
as the masses and 
mixings above produce a transition probability too small to be
of significance \cite{PET}.

The other case, $X^{[\nu]}=0$, yields more massive neutrinos. In 
this case, we have a supersymmetric zero in the Dirac mass 
matrix, leading to 
\be 
\widehat Y^{[\nu]}\sim{v_u^2\over M\lambda_{c}^{}}
\left( \begin{array}{ccc}
1&\lambda_c^{3} &\lambda_c^3\\ 
\lambda_c^{3}&\lambda_c^{5}&\lambda_c^{5}\\
\lambda_c^3&\lambda_c^{5}&\lambda_c^3\end{array}  \right)\ .\ee
We obtain the same MNS lepton mixing matrix as in the previous 
case, produce the same inverted neutrino mass hierarchy, but 
obtain different orders of magnitude estimates for the neutrino 
masses
\be
m_{\nu_e}\sim {v_u^2\over M\lambda_{c}^{}}\sim 5\times 
10^{-3,-4}~{\rm eV}\ ;\qquad
m_{\nu_\mu}\sim m_{\nu_\tau}\sim 10^{-6,-7}~~{\rm eV}\ .\ee
Although the magnitudes of $\Delta m^2$ and the mixing angle
are naively consistent with the small angle MSW solution to the solar 
neutrino crisis in this case, the inverted neutrino hierarchy unfortunately
prohibits the resonance condition from being satisfied \cite{PET} \cite{SMI},
as it produces a $\Delta m^2$ of the incorrect sign.  It is interesting to
note, however, that the resonant {\it anti}-neutrino oscillation that 
results from this sign is consistent with the observed anti-neutrino
flux from supernova SN1987A \cite{SMI}.

The most striking result of this analysis is the 
inverted 
hierarchy among neutrino masses mentioned above. Different choices for the 
$\overline N$ charges change neither this hierarchy nor the 
nature of the mixing. This is because the order of magnitude 
matrices factorize, and the $\overline N$ charge 
contributions to the exponents therefore cancel in the seesaw matrix. Thus, 
in our model with three right-handed neutrinos, the inverted hierarchy 
is a consequence of the 
charge assignment of the charged leptons!  An important exception
to this rule, however, arises when there are sufficient supersymmetric
zeroes in the matrix $Y^{[0]}$ for it to develop a zero eigenvalue.
In this interesting case, one of the neutrinos will have vanishing
Majorana mass, will drop out of the seesaw mechanism, and will 
therefore obtain a mass of the order of the electroweak scale, possibly
suppressed by some powers of $\lambda$.  This allows for the possibility
of a normal hierarchy, and potential agreement with the MSW 
requirements.

As an example, consider the situation in which the $\overline N_i$
fields are assigned $X$ charges $-{1 \over 2}, -{1 \over 2}$, and
$-{3 \over 2}$, and $Y^{(2)}$ charges 1, 0, and $-1$, respectively.  The
tau neutrino then drops out of the seesaw, and the electron and 
muon neutrinos seesaw with $\Delta m^2$ and $\sin^2 2\Theta$ parameters
consistent with the MSW effect.  The tau neutrino is suppressed from
the electroweak scale by 5 powers of $\lambda$, and so picks up a 
Dirac mass around 50 MeV.
The masses and mixings of the standard model particles are of 
course exactly the same as those obtained for the inverted hierarchy
examples above, but the S-field sector will be somewhat modified
so as to cancel the required anomalies.
The mass of the tau neutrino is order of magnitude consistent
with the current experimental upper limit of 24 MeV \cite{PDG},
but in potential conflict with cosmology.
The tau neutrino is unstable, however, decaying preferentially to 
$\nu_{\mu} \ \gamma$,
and if it does so sufficiently rapidly, it can evade the problem
of overclosure of the universe.
Although this solution may
be phenomenologically viable, it is not as generic as the solutions 
discussed earlier containing the inverted hierarchy.  For 
this reason, we comment only briefly on it here, and leave a comprehensive
study of models with zero eigenvalues in $Y^{[0]}$ to a future
publication.

One of us would like to acknowledge the hospitality of the Rutgers
particle theory group where this paper was completed.


\begin{thebibliography}{Ref}
\bibitem{SMI} Thanks to Prof. A. Smirnov for calling our attention 
to these facts.
\end{thebibliography}
\end{document}